# Tensors in Power System Computation I: Distributed Computation for Optimal Power Flow, DC OPF

HyungSeon Oh, *Member, IEEE*

*Abstract –* Tensor decomposition plays a key role in identifying common features across a collection of matrices in many areas of science. A fundamental need in big data research is to process data tabulated as large-scale matrices using eigenvectors. A higher-order generalized singular value decomposition technique successfully captures the common features of the same organ from multiple animals in genomic signal processing. A recent SDP (semi-definite programming) approach to solve an AC optimal power flow was accompanied by the problem formulation in the Cartesian coordinate system. The collection of nodal Kirchhoff's laws introduces a 3D tensor with a common feature of individual matrices to maintain local power balance. In this paper, the mathematical process is established and the common feature is identified. The common feature is a key element to a fully decentralized and therefore scalable algorithm to solve AC optimal power flow.

## I. INTRODUCTION

COMPARATIVE analyses of large-scale data explores the characteristic features of the data. From the comparative analyses of global mRNA expression from multiple model organisms, the fundamental understanding of the universality in the molecular biological mechanisms is enhanced [1]. The conventional framework is the generalized singular value decomposition (GSVD) [2-5] that is limited to two matrices. Recent study in higher-order generalized singular value decomposition (HOGSVD) provides a mathematical framework in comparing the same organ from multiple animals and in extracting a common feature from them [6, 7]. In this work, Ponnapalli advanced the idea of an HO GSVD in which each matrix is decomposed into a product of the form $A_k = U_k \Sigma_k W^T$ where $A_k$ is a representative matrix related to the organ from the k$^{th}$ animal. The right-hand-side eigenvector $W$ captures the common feature of the organ from all the animals considered under the study.

Owing to the interest of the global optimizer, the AC optimal power flow problem is formulated in the Cartesian coordinate system [8-11]. In the formulation, each equation includes a matrix indicating how the global voltage vector affects the injection of power at each node, i.e., $v^T \Phi_j^p v = p_j$ where $v$, $\Phi_j^p$, and $p_j$ are voltage, a matrix associated with Node $j$, and the real power injection at the node, respectively. Even though the impacts can be different, the individual matrices represent the impact of the voltage vector on the nodal injection. A similarity exists between the same organ from multiple animals and a nodal impact $\Phi_j^p$ of power injection at multiple locations. Therefore, it is possible to extract a common feature among matrices.

Related to the matrices, the power balance equations are affected by global and local features. The extraction of global features from local matrices is the key component of our distributed algorithm with the computation and dispatch decisions localized at a node in the power network. The OPF problem is completely solved by individual nodes, which have limited information from the system. This distributed computation approach has been pursued in considering the parallel computation utilizing matrix factorization in centralized power flow algorithms [12-15]. Baldick et al. [16, 17] proposed a regional decomposition approach that solves a large-scale power problem by dividing it into small, multiple problems associated with a few overlapping subsystems so that each subsystem's problem is solved in parallel. Similar ideas are further developed to solve multi-area coordination problems in an interconnected market environment [18, 19]. Some distributed algorithms leverage methods of multipliers based on an alternating direction method of multiplier (ADMM, [20]) [21], but they ignore highly nonlinear flow constraints. Furthermore, the algorithm may not be scalable if some nodes are heavily connected, and the computation costs amongst sub-problems are not evenly distributed according to the local connectivity at each node. For a radially connected network, the distributed OPF sub-problems have only four variables of which computation cost is low [22-24].

Numerous advantages exist for our proposed algorithm for solving the AC OPF problem for a mesh network: 1) A closed form solution is found for each optimization sub-problem, thus eliminating the need for an iterative procedure for each ADMM iteration; 2) only four variables are considered for each sub-problem to make the algorithm scalable with respect to the size of the mesh network; and 3) there exists great similarity among sub-problems. The rest of the paper is structured as follows. The mathematical derivations to the tensor decomposition and to the ADMM sub-problems are outlined in section II. In section III, we

---

H. Oh is with the State University of New York at Buffalo (e-mail: hso1@buffalo.edu).

develop our distributed algorithm. In section IV, we test its scalability using data from the IEEE model system and a real-world network. We conclude this paper in section V.

## II. TENSOR FORMULATION

The matrices associated at the $j^{th}$ node appearing the Kirchhoff's laws in balancing the real ($\Phi_j^p$) and the reactive ($\Phi_j^q$) power injections are:

$$\Phi_j^p = \eta_j^p \left( e_j e_j^T + e_{j+N} e_{j+N}^T \right) + \mu_j \left[ \left( \xi_j^T \ \zeta_j^T \right)^T e_j^T \right. \tag{1}$$
$$\left. + e_j \left( \xi_j^T \ \zeta_j^T \right) + \left( -\zeta_j^T \ \xi_j^T \right)^T e_{j+N}^T + e_{j+N} \left( -\zeta_j^T \ \xi_j^T \right) \right]$$

$$\Phi_j^q = \eta_j^q \left( e_j e_j^T + e_{j+N} e_{j+N}^T \right) + \mu_j \left[ \left( \zeta_j^T \ -\xi_j^T \right)^T e_j^T \right. \tag{2}$$
$$\left. + e_j \left( \zeta_j^T \ -\xi_j^T \right) + \left( \xi_j^T \ \zeta_j^T \right) e_{j+N}^T + e_{j+N} \left( \xi_j^T \ \zeta_j^T \right) \right]$$

where $\eta$ and $\mu$ are scalar; $\xi$ and $\zeta$ are vectors with cardinality of $N$ (the number of nodes), and their $j^{th}$ elements are zeros; $e_j$ is the $j^{th}$ column vector of an identity matrix with the cardinality of twice the number of nodes.

Equation (1) is simplified as follows:

$$\Phi_j^p = \eta_j^p \left( e_j e_j^T + e_{j+N} e_{j+N}^T \right) + \mu_j \phi_j II \phi_j^T$$

$$\text{where } \phi_j = \frac{1}{\sqrt{2}} \begin{bmatrix} \xi_j + e_j & \xi_j - e_j & -\zeta_j & -\zeta_j \\ \zeta_j & \zeta_j & \xi_j + e_j & \xi_j - e_j \end{bmatrix} \tag{3}$$

$$\text{and } II = diag\left[\begin{pmatrix} 1 & -1 & 1 & -1 \end{pmatrix}\right]$$

The matrix $\phi_j$ is a full column-rank matrix and each column vector is orthonormal with each other. Suppose $Q_2$ with the cardinality of $2N$-by-$(2N-4)$ spans the null space of $\phi_j$. Then, one finds

$$Q_2^T \phi_j = 0 \rightarrow Q_2^T \begin{pmatrix} \xi_j \\ \zeta_j \end{pmatrix} = 0, Q_2^T \begin{pmatrix} -\zeta_j \\ \xi_j \end{pmatrix} = 0, Q_2^T e_j = 0, Q_2^T e_{j+N} = 0 \tag{4}$$

Similar to Eq. (3), one finds

$$\Phi_j^q = \eta_j^q \left( e_j e_j^T + e_{j+N} e_{j+N}^T \right) + \mu_j \bar{\phi}_j II \bar{\phi}_j^T$$

$$\text{where } \bar{\phi}_j = \frac{1}{\sqrt{2}} \begin{bmatrix} \zeta_j + e_j & \zeta_j - e_j & \xi_j & \xi_j \\ -\xi_j & -\xi_j & \zeta_j + e_j & \zeta_j - e_j \end{bmatrix} \tag{5}$$

From Eq. (4), it is easy to see that $Q_2$ also spans the null space of $\bar{\phi}_j$, i.e. $\phi_j$ spans the same space, and they share the same null space. It is also clear that the first elements in Eqs. (1) and (2) are orthogonal to $Q_2$ because $Q_2^T e_j = 0$ and $Q_2^T e_{j+N} = 0$. The eigenvalue decomposition of $\Phi_j^p$ becomes:

$$\Phi_j^p = \phi_j \left\{ \mu_j \left[ \varsigma_{j+}^p \left( \varsigma_{j+}^p \right)^T + \varsigma_{j-}^p \left( \varsigma_{j-}^p \right)^T + II \right]^{4 \times 4} \right\} \phi_j^T$$

$$\text{where } \begin{cases} \varsigma_{j+}^p = \sqrt{\dfrac{\eta_j^p}{\mu_j}} \phi_j^T e_j = \sqrt{\dfrac{\eta_j^p}{\mu_j}} \begin{pmatrix} 1 & -1 & 0 & 0 \end{pmatrix}^T \\ \varsigma_{j-}^p = \sqrt{\dfrac{\eta_j^p}{\mu_j}} \phi_j^T e_{j+N} = \sqrt{\dfrac{\eta_j^p}{\mu_j}} \begin{pmatrix} 0 & 0 & 1 & -1 \end{pmatrix}^T \end{cases} \tag{6}$$

The matrix in the curly bracket is the rank-2 update of the $II$ matrix. The real space spanned by $\Phi_j^p$ is the same as $\phi_j$, and therefore $Q_2$ spans the null space of $\Phi_j^p$.

$$\Phi_j^p = \frac{\mu_j}{2} \tilde{\phi}_j II \tilde{\phi}_j^T = B_{j,p} II B_{j,p}^T \tag{7}$$

where

$$\tilde{\phi}_j = \begin{bmatrix} \xi_j + K_{j+}^p e_j & \xi_j - K_{j-}^p e_j & -\zeta_j & -\zeta_j \\ \zeta_j & \zeta_j & \xi_j + K_{j+}^p e_j & \xi_j - K_{j-}^p e_j \end{bmatrix}$$

$$K_{j+}^p = 1 + \frac{\eta_j^p}{2\mu_j}, \ K_{j-}^p = 1 - \frac{\eta_j^p}{2\mu_j}, \ \text{and } B_{j,p} = \sqrt{\frac{\mu_j}{2}} \tilde{\phi}_j$$

The collection of the matrix $B_{j,p}$ is a 3D array of elements, i.e., a 3D tensor that indicates how voltage affects the real and the reactive power injections. The cardinality of each $B_{j,p}$ is $2N$-by-$4$. The singular value decomposition of the 2D collection of $B_j$ yields:

$$B_j = V U_j \Sigma_j W^T \tag{8}$$

where $\Sigma_j$ is a diagonal matrix. HOGSVD finds the common right-hand-side nonorthogonal eigenvector $W$.

As a result,

$$v^T \Phi_j^p v = v^T V U_j \Sigma_j \left( W^T II W \right) \Sigma_j U_j^T V^T v \tag{9}$$

where voltage $v$ lies in the space of $V^T$, a column vector of $V$. By defining $\alpha$ and $\beta$ for real and reactive power injection, respectively, i.e., $\alpha_j = \Sigma_j U_j^T V^T v \ \forall j \leq N$, and $\beta_j = \Sigma_{j+N} U_{j+N}^T V^T v \ \forall N+1 \leq j \leq 2N$.

The QR-factorization of $U_j \Sigma_j = \begin{bmatrix} Q_{R,j} & Q_{N,j} \end{bmatrix} \begin{bmatrix} R_j^T & 0^T \end{bmatrix}^T$ yields $\Sigma_j U_j^T = R_j^T Q_{R,j}^T$. Because $\Phi_j^p$ and $\Phi_j^q$ span the same space and $V$ and $W$ are full-rank matrices, $\Sigma_{j+N} U_{j+N}^T$ spans the same space of $\Sigma_j U_j^T$. Therefore, $\alpha$ and $\beta$ are linearly dependent:

$$\beta_j = \Sigma_{j+N} U_{j+N}^T V^T v = \left( R_{j+N}^T R_j^{-T} \right) \alpha_j \tag{10}$$

The power balance equations become:

$$v^T \Phi_j^p v = \alpha_j^T \left( W^T II W \right) \alpha_j = \alpha_j^T III \alpha_j = p_j^{inj}$$
$$v^T \Phi_j^q v = \beta_j^T \left( W^T II W \right) \beta_j = \alpha_j^T III_j \alpha_j = q_j^{inj} \tag{11}$$

where $III = W^T II W$ and $III_j = R_j^{-1} R_{j+N} III R_{j+N}^T R_j^{-T}$. Note that the matrix $III$ is identical for all the nodes.

Let $\gamma_j$ be a complimentary vector of $\alpha$, because $Q_{j,N}$ spans the null space of $U_j \Sigma_j$,

$$\begin{pmatrix} \alpha_j \\ \gamma_j \end{pmatrix} = \begin{bmatrix} R_j^T & 0 \\ 0 & I \end{bmatrix} \begin{pmatrix} Q_{R,j}^T \\ Q_{N,j}^T \end{pmatrix} V^T v \rightarrow \begin{cases} \alpha_j = R_j^T Q_{R,j}^T V^T v \\ \gamma_j = Q_{N,j}^T V^T v \end{cases} \tag{12}$$

Note that HOGSVD yields non-orthogonal $U_j$ and $W$ matrices, and therefore $U_{j+N}^T U_j$ is not zero. The voltage also affects power flows over the grid, and the impacts are expressed in matrices. For a line $k$ connecting the injection node $j$ and the ejection nodes $i$, the matrices for the real and reactive power flows over the line are:

$$\Psi_{k,j}^p = \eta_{k,j}^p \left( e_j e_j^T + e_{j+N} e_{j+N}^T \right) + \mu_{k,j} \left[ \left( \xi_{k,j}^T \ \zeta_{k,j}^T \right)^T e_j^T \right. \tag{13}$$
$$\left. + e_j \left( \xi_{k,j}^T \ \zeta_{k,j}^T \right) + \left( -\zeta_{k,j}^T \ \xi_{k,j}^T \right)^T e_{j+N}^T + e_{j+N} \left( -\zeta_{k,j}^T \ \xi_{k,j}^T \right) \right]$$

$$\Psi_{k,j}^{q} = \eta_{k,j}^{q}\left(e_{j}e_{j}^{T} + e_{j+N}e_{j+N}^{T}\right) + \mu_{k,j}\left[\left(\zeta_{k,j}^{T} - \xi_{k,j}^{T}\right)^{T}e_{j}^{T}\right.$$
$$\left. + e_{j}\left(\zeta_{k,j}^{T} - \xi_{k,j}^{T}\right) + \left(\xi_{k,j}^{T}\zeta_{k,j}^{T}\right)^{T}e_{j+N}^{T} + e_{j+N}\left(\xi_{k,j}^{T}\zeta_{k,j}^{T}\right)\right] \quad (14)$$

A bus can be connected by multiple lines, which leads to $\xi_j = \sum_{k\in\Omega_j}\xi_{k,j}$ and $\zeta_j = \sum_{k\in\Omega_j}\zeta_{k,j}$ where $\Omega_j$ is the set of lines connected to node $j$. Similar to Eq. (3), (14) becomes:

$$\Psi_{k,j}^{p} = \eta_{k,j}^{p}\left(e_{j}e_{j}^{T} + e_{j+N}e_{j+N}^{T}\right) + \mu_{k,j}\psi_{j}II\psi_{j}^{T} = K_{k,j}S_{k,j}K_{k,j}^{T}$$

$$\text{where } \psi_j = \frac{1}{\sqrt{2}}\begin{bmatrix} \xi_{k,j}+e_j & \xi_{k,j}-e_j & -\zeta_{k,j} & -\zeta_{k,j} \\ \zeta_{k,j} & \zeta_{k,j} & \xi_{k,j}+e_j & \xi_{k,j}-e_j \end{bmatrix} \quad (15)$$

$K$ and $S$ are eigenvector and eigenvalue pairs. It is noted that $\phi_j^T\psi_j \neq 0$, and $\phi_j$ is a full-column rank matrix. Therefore, $\psi_j$ lies in the space spanned by $\phi_j$, i.e., $\psi_j$ is orthogonal to $Q_{j,N}$, and $\Phi_{k,j}^p$ and $\Psi_{k,j}^p$ share the null space that are revealed from the eigenvalue decomposition. The similarity transformation with $V$ makes $V\Phi_{k,j}^p V^T$ and $V\Psi_{k,j}^p V^T$ share the null space also. The real and the reactive power flows over the like $k$ connecting node $j$ are:

$$v^T\Psi_{k,j}^p v = \alpha_j^T\mathcal{F}_{k,j}^p\alpha_j + 2\alpha_j^T\mathcal{G}_{k,j}^p\xi_{k,j} + \xi_{k,j}^T S_{k,j}\xi_{k,j} \quad (16)$$

where $\mathcal{F}_{k,j}^p = R_j^{-1}Q_{Rj}^T V^T\Psi_{k,j}^p VQ_{Rj}R_j^{-T}$, $\mathcal{G}_{k,j}^p = R_j^{-1}Q_{Rj}^T V^T S_{k,j}$

and $\xi_{k,j} = K_{k,j}^T VQ_{Nj}Q_{N,j}^T V^T v$

Because $\Psi_{k,j}^p$ and $\Psi_{k,j}^q$ share the same real and null spaces, their similarity transformation $V^T\Phi_{k,j}^p V$ and $V^T\Psi_{k,j}^q V$ share the same spaces. Therefore, the reactive power flow is expressed in terms of $\alpha_j$ and $\xi_{k,j}$:

$$v^T\Psi_{k,j}^q v = \alpha_j^T\mathcal{F}_{k,j}^q\alpha_j + 2\alpha_j^T\mathcal{G}_{k,j}^q\xi_{k,j} + \xi_{k,j}^T \mathcal{I}_{k,j}^q\xi_{k,j} \quad (17)$$

where $\mathcal{F}_{k,j}^q = R_j^{-1}Q_{Rj}^T V^T\Psi_{k,j}^q VQ_{Rj}R_j^{-T}$, $\mathcal{G}_{k,j}^q = R_j^{-1}Q_{Rj}^T V^T\Psi_{k,j}^q K_{k,j}$, and $\mathcal{I}_{k,j}^q = K_{k,j}^T\Psi_{k,j}^q K_{k,j}$, a similarity transformation of $\Psi_{k,j}^q$.

Eqs. (16) and (17) show that the real and the reactive power flow over a line $k$ are quadratic with $\alpha_j$ and $\xi_{k,j}$.

The voltage constraints are also associated with matrices:

$$\underline{|v_j|}^2 \leq v^T\left(e_j e_j^T + e_{j+N}e_{j+N}^T\right)v \leq \overline{|v_j|}^2 \quad (18)$$

The matrix inside the parenthesis is rank-2, and $(e_j+e_{j+N})^T V$ and let $\delta_j$ be $E_j^T V$ where $E_j = [e_j\ e_{j+N}]$.

$$v^T\left(e_j e_j^T + e_{j+N}e_{j+N}^T\right)v = v^T E_j E_j^T v = \delta_j^T\delta_j \quad (19)$$

As a result, $\alpha_j$, $\delta_j$, and $\xi_{k,j}$ are the nodal and low-dimensional projections $x_j$ of the global voltage vector, and they are in the cardinality of 4, 2, and 4, respectively.

## III. DISTRIBUTED COMPUTATION FOR OPF, DC OPF

ADMM achieves the superior convergence properties of the method of multipliers from the decomposability of dual decomposition [20]. It intends to solve an optimization problem with a trust-region sized of $\Delta_j$:

$$\min_{x_j,v} \sum_{j=1}^{N}\left[f_j(x_j) + \frac{\mu_j}{2}\left(\|x_j - M_j v\|_2^2 - \Delta_j^2\right)\right] + g(v) \quad (20)$$

The augmented Lagrangian becomes:

$$\mathcal{L}(x_j,v) = \sum_{j=1}^{N}\left[f_j(x_j) + \frac{\mu_j}{2}\left(\|x_j - M_j v\|_2^2 - \Delta_j^2\right)\right] + g(v) \quad (21)$$

The following steps are taken at the $m^{th}$ iteration:

$$x_j^m \leftarrow \operatorname*{argmin}_{x_j}\mathcal{L}\left(x_j, v^{m-1}\right) \forall j;\ v^m \leftarrow \operatorname*{argmin}_{x}\mathcal{L}\left(x^m, v\right) \quad (22)$$

An OPF problem is formulated:

$$\min_{v,p,q} C(p)$$

$$s.t.\begin{cases} v^T\Phi_j^p v = p_j^g - p_j^d & \forall j \\ v^T\Phi_j^q v = q_j^g - q_j^d & \forall j \\ \underline{|v_j|}^2 \leq v^T\left(e_j e_j^T + e_{j+N}e_{j+N}^T\right)v \leq \overline{|v_j|}^2 & \forall j \\ \left(v^T\Psi_{k,j}^p v\right)^2 + \left(v^T\Psi_{k,j}^q v\right)^2 \leq cap_k^2 & \forall k \\ \left(\underline{p}^T\ \underline{q}^T\right)^T \leq \left(p^T\ q^T\right) \leq \left(\overline{p}^T\ \overline{q}^T\right)^T \end{cases} \quad (23)$$

Note that the flow constraint is imposed on the injection-side as well as on the ejection-side of the line $k$. This OPF problem is reformulated with the tensor decomposition developed in Section II.

For separating a nordal variable at the $j^{th}$ node and local interactions, we introduce $\alpha_j$ to formulate an optimization problem as follows:

$$\min_{x_j,v,p_j,q_j} \sum_j C_j(p_j)$$

$$s.t.\begin{cases} \alpha_j^T III_j\alpha_j = p_j - p_j^d \text{ and } \alpha_j^T III_j\alpha_j = q_j - q_j^d & \forall j \\ \left(\alpha_j^T\mathcal{F}_{k,j}^p\alpha_j + 2\alpha_j^T\mathcal{G}_{k,j}^p\xi_{k,j} + \xi_{k,j}^T S_{k,j}\xi_{k,j}\right)^2 \\ + \left(\alpha_j^T\mathcal{F}_{k,j}^q\alpha_j + 2\alpha_j^T\mathcal{G}_{k,j}^q\xi_{k,j} + \xi_{k,j}^T\mathcal{I}_{k,j}^q\xi_{k,j}\right)^2 \leq cap_k^2\ \forall k \\ \underline{|v_j|}^2 \leq \delta_j^T\delta_j \leq \overline{|v_j|}^2 & \forall j \\ \underline{p}_j \leq p_j \leq \overline{p}_j, \underline{q}_j \leq q_j \leq \overline{q}_j & \forall j \\ x_j = \begin{pmatrix}\alpha_j\\\delta_j\\\xi_{k,j}\end{pmatrix} = \begin{bmatrix}R_j^T Q_{R,j}^T V^T\\ E_j^T\\ K_{k,j}^T VQ_{Nj}Q_{N,j}^T V^T\end{bmatrix}v = M_j v & \forall j \end{cases} \quad (24)$$

where $\underline{p}_j$, $\overline{p}_j$, $\underline{q}_j$, and $\overline{q}_j$ are set to zero for all PQ buses. The last three equality constraints are termed "consensus constraints". The augmented Lagrangian can be completely decomposed to subproblems at each node.

At a PV bus where a generator is located, a distributed optimization problem is formulated by relaxing the consensus equality constraints and by including the constraints into the objective function. The equality constraints are removed by replacing the real and the imaginary power generations out of the generator located at the bus. Therefore, the distributed optimization problem becomes:

$$\min_{x_j} C_j(\alpha_j) + \frac{\mu_j}{2}\left(\left\|x_j - M_j v^m\right\|_2^2 - \Delta_j^2\right): \quad s.t.$$

$$\begin{cases}
\left(\alpha_j^T \mathcal{F}_{k,j}^p \alpha_j + 2\alpha_j^T \mathcal{G}_{k,j}^p \xi_{k,j} + \xi_{k,j}^T S_{k,j} \xi_{k,j}\right)^2 \\
+ \left(\alpha_j^T \mathcal{F}_{k,j}^q \alpha_j + 2\alpha_j^T \mathcal{G}_{k,j}^q \xi_{k,j} + \xi_{k,j}^T \mathcal{T}_{k,j}^q \xi_{k,j}\right)^2 \leq cap_k^2 \; k \in \Omega_j \\
\left|\underline{v_j}\right|^2 \leq \delta_j^T \delta_j \leq \left|\overline{v_j}\right|^2 \\
\underline{p_j} - p_j^d \leq \alpha_j^T III \alpha_j \leq \overline{p_j} - p_j^d \\
\underline{q_j} - q_j^d \leq \alpha_j^T III_j \alpha_j \leq \overline{q_j} - q_j^d
\end{cases} \quad (25)$$

By setting a large value for $\mu_j$, the consensus constraints are restricted within a trust-region sized of $\Delta_j$, and (25) becomes a convex optimization problem. For a sufficiently large value of $m$, $\Delta_j$ approaches zero so that the consensus constraints recover the definition of the nodal projection $x_j$.

Similarly, for a PQ bus, the distributed optimization problem becomes:

$$\min_{x_j} N_p\left(\alpha_j^T III \alpha_j + p_j^d\right)^2 + N_q\left(\alpha_j^T III_j \alpha_j + q_j^d\right)^2$$
$$+ \frac{\mu_j}{2}\left(\left\|x_j - M_j v^m\right\|_2^2 - \Delta_j^2\right): \quad s.t. \quad (26)$$

$$\begin{cases}
\left(\alpha_j^T \mathcal{F}_{k,j}^p \alpha_j + 2\alpha_j^T \mathcal{G}_{k,j}^p \xi_{k,j} + \xi_{k,j}^T S_{k,j} \xi_{k,j}\right)^2 \\
+ \left(\alpha_j^T \mathcal{F}_{k,j}^q \alpha_j + 2\alpha_j^T \mathcal{G}_{k,j}^q \xi_{k,j} + \xi_{k,j}^T \mathcal{T}_{k,j}^q \xi_{k,j}\right)^2 \leq cap_k^2 \; k \in \Omega_j \\
\left|\underline{v_j}\right|^2 \leq \delta_j^T \delta_j \leq \left|\overline{v_j}\right|^2
\end{cases}$$

For sufficiently large values of $N_p$ and $N_q$, the power balance equations are forced to meet, and $\mu_j$ needs to be very large for making (26) a convex problem.

With the update of the local projections $x_j$ described in (25) and (26), an optimization problem to update the global variable $v$ is formulated as follows:

$$\min_v \sum_{j=1}^N \frac{\mu_j}{2}\left(\left\|x_j^m - M_j^T v\right\|_2^2 - \Delta_j^2\right) \; s.t. \; e_{ref+N}^T v = 0 \quad (27)$$

Eq. (27) is a convex optimization problem with a single equality constraint, and one finds an analytic solution to the problem:

$$v^m = \sum_{j=1}^N v_j^m \quad \text{where } v_j^m = \omega_j - \frac{e_{ref+N}^T \omega_j}{e_{ref+N}^T \tau_j} \tau_j \quad (28)$$

where $\omega_j = \left(\mathcal{M}^{-1} \mu_j M_j\right) x_j^m$, $\tau_j = \mathcal{M}^{-1} e_{ref+N}$, and $\mathcal{M} = \sum_{j=1}^N \left(\mu_j M_j M_j^T\right)$. Note that the matrices $M_j$ and $\mathcal{M}$ are fixed at iteration, and therefore it is pre-computed and supplied, and therefore, $v_j^m$ can be computed once $x_j^m$ is identified from (25) and (26). If all $\mu$'s are of the same value, then $v_j^m$ does not depend on the value of $\mu$. The computation cost associated with (28) is linear with $N$, i.e., $\vartheta(N)$. (27) is the only optimization problem where the control variable is in a high dimension, but the analytical solution exists, and furthermore, $v_j^m$ is computed directly from $x_j^m$. Therefore, this DC OPF framework yields a scalable algorithm to solve an OPF problem without any approximations and/or any assumptions.

## IV. NUMERICAL RESULTS & DISCUSSIONS

The tensor decomposition outlined in Section II is performed for IEEE model systems, and the common right-hand side orthonormal matrix $V$ on the IEEE 4-bus cases is illustrated in Figure 1 (left figure). Because $V$ is a full column matrix, the voltage vector $v$ spans the column space of $V$. If $v$ satisfies a power flow case, the rotated $v$ in the polar coordinate system does too. This symmetry imposed on $v$ is also reflected in the column space in $V$. Suppose a column vector in $V$ is in the form of $[v_x; v_y]$, due to orthogonality of $V$, there is a column vector of $[-v_y; v_x]$. The pattern is found in Figure 1. A majority of the column vectors in $V$ is sparse if small numbers are ignored such as $10^{-5}$. Figure 1 also illustrates the common right-hand side matrix $W$ obtained from the tensor decomposition described in Section II. Even if $W$ is not orthogonal in general, $W$ for all the IEEE model systems happens orthonormal, and furthermore, they satisfy $W^T II W = WIIW^T = -II$, i.e., $III = -II$.

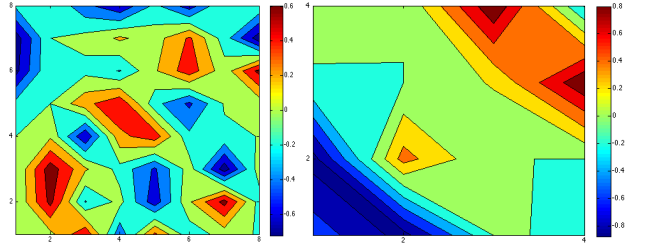

Figure 1. The common left-side eigenvector $V$ (left) and right-hand side eigenvector $W$ (right) from 4-bus system.

For comparison, we implemented an ADMM approach outlined in Ref. [21]. We were not able to find a converging solution for a mesh network 3-bus case. We modified the formulation so that the distributed optimization problem at a PQ bus are different from that at a PV bus, i.e., the power balance equation comes in the objective function at PQ buses. With the change, we were able to find a converging case. Note that the formulation in Ref. [21] does not include the flow constraints, and nor does ours. Figure 2 shows the result of the approach. After 27 iterations, ADMM finds a feasible solution to OPF at the 3-bus system. Even though the solution satisfies the power balance equation and other constraints, it was observed that the power flow constraints were violated if imposed. For a system with more than 3-buses, this ADMM approach fails to find a solution. We checked the results, and found that, different from a radial network, an estimate of a nodal voltage could be conflicting and the local exchange of voltage information with neighbors in this ADMM approach was not sufficient to adjust conflicting information.

Different from the ADMM approach, the proposed DC OPF optimizes a nodal projection $x_j$ of the global voltage variable $v$. As a result, much less conflicts exist and no local exchange of information is necessary among neighbors. Even if there exist conflicts, they vanish rapidly as iteration proceeds. For the 3-bus system, the proposed DC OPF algorithm is formulated, including flow constraints. For a faster convergence, social responsibility constraints are imposed. The social responsibility constraints refer that if a generator is located inside a load pocket, it needs to generate at least local loads when the tie-lines do not have a sufficient capacity to bring power into the load pocket, which effectively increases the minimum power generation constraint. Figures 3 (A), (B), and (C) graphically present the DC OPF results for 3-, 4-, and 30-bus systems, respectively. All the cases are modified so that at least one transmission line is congested, and DC OPF finds the numerically same solutions as ones with MATPOWER [26]. [27].

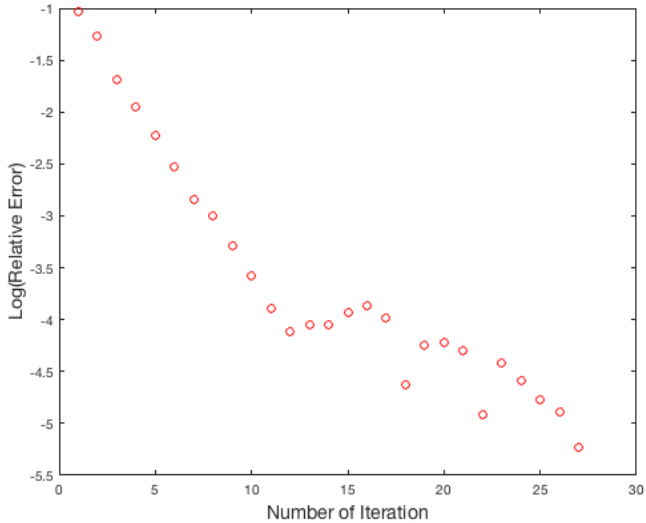

Figure 2. The performance of the ADMM approach for a 3-bus case where relative error refers to the change in voltage at iteration.

Because the computation time to solve a sub-optimization problem in (25) and (26) and the number of iterations to converge a solution do not increase with the system size, the computation time solving (25) and (26) stay unchanged with respect to the system size if enough number of cores is provided for computation. However, the computation process in (28) increases linearly with the size of the system. Therefore, the computation cost of this DC OPF algorithm is in $\vartheta(N)$. Therefore, this DC OPF is a scalable algorithm for a large-scale network. At this point, DC OPF is not yet implemented in parallel computation. Future works include the parallel computation to solve an OPF problem for a large-scale network.

For the ADMM approach, this condition was met for 3-bus case because a 3-bus mesh network allows each bus to access the entire voltage vector. This visibility is provided for a completely connected system, i.e., all the nodes are connected with each other. However, for a system with 4 or more buses, mesh networks are not completely connected. Different from the ADMM approach, for the 3-bus case (Figure 3 (A)), in early stages of the iteration process, there is conflict in voltage information, but the conflict is quickly resolved among the information exchange outlined in (28) – see the spike around the 10[th] period. The proposed DC OPF framework comprises 1) optimizing a nodal projection of voltage vector locally, and 2) recovering the voltage vector from local projections. Any conflict will be resolved within a small number of iterations. This makes it possible for DC OPF to converge.

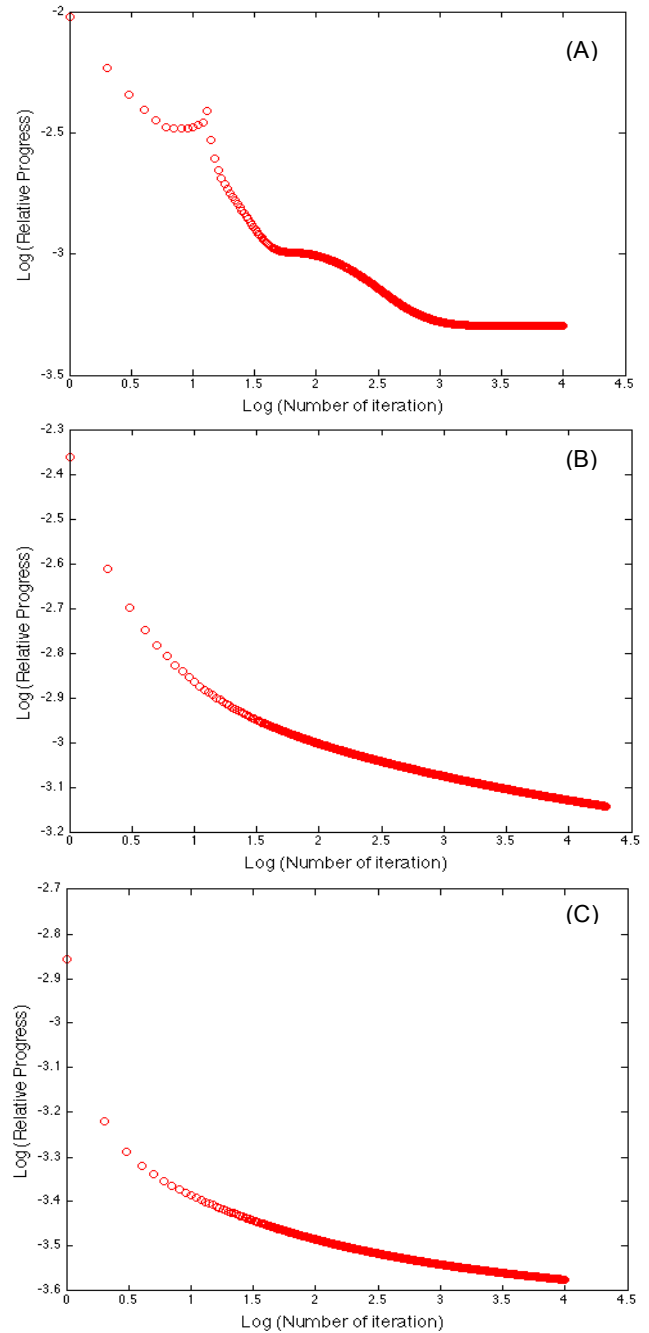

Figure 3. The performance of DC OPF for (A) a 3-bus case, (B) a 4-bus case, and (C) a 30-bus case. All the cases include at least one congested line.

## V. CONCLUSIONS

Tensor-based computation becomes a key component in modern numerical computation. It shows great features to extract common properties from discrete samples, and to quantify the impact of each eigenvector. A drawback of tensor-based computation is the lack of structural properties of resulting eigenvectors such as orthogonality. Inherited from the symmetry in power systems, orthogonality of both left-hand and right-hand common eigenvectors is observed in all the IEEE model systems. This common eigenvectors make it possible to project the voltage vector on to low dimensional nodal control variables. A least-square approach helps to reconstruct the voltage vector from the nodal control variables. This projection-reconstruction approach is integrated to solve an optimal power flow problem.

AC OPF is an NP-hard problem, the solution to which incurs a high computation cost for a large-scale power system. Various heuristic approaches have been proposed for solving AC OPF, but no markets—at least none in the United States—are operated based on AC OPF. A distributed computation for optimal power flow (DC OPF) is proposed to solve AC OPF without any approximations and/or assumptions. The computational performance is tested on several meshed IEEE model systems. Different from other distributed computation methods that focus on the problem with radial networks, this proposed DC OPF successfully solves the optimal power flow problems for mesh networks. Implemented in parallel computation, this DC OPF algorithm can make it possible to utilize AC OPF for a large-scale power system operation.